\documentclass[conference]{IEEEtran}
\IEEEoverridecommandlockouts% chktex 1

\usepackage{cite}
\usepackage{amsmath,amssymb,amsfonts}
\usepackage{graphicx}
\usepackage{textcomp}
\usepackage{xcolor}
\usepackage{array}
\usepackage{tabularx}
\usepackage{booktabs}
\newcolumntype{Y}{>{\raggedright\arraybackslash}X}
\usepackage{multirow}
\usepackage{stfloats}
\usepackage{tikz}
\usepackage{url}
\usepackage{microtype}
\usetikzlibrary{arrows.meta,positioning}

\begin{document}
\bstctlcite{IEEEexample:BSTcontrol}

\title{Adaptive Perturbation Selection for Contrastive Audio Decoding}

\author{
\IEEEauthorblockN{Aaron Isidore Grace}
\IEEEauthorblockA{
  \textit{Department of Computer Science} \\
  \textit{University of Iowa} \\
  Iowa City, IA, USA \\
  aaron.grace@uwaterloo.ca
}
\and
\IEEEauthorblockN{Zhouyuan Huo}
\IEEEauthorblockA{
  \textit{Google} \\
  Mountain View, CA, USA \\
  huozhouyuan@gmail.com
}
\and
\IEEEauthorblockN{Weiran Wang}
\IEEEauthorblockA{
  \textit{Department of Computer Science} \\
  \textit{University of Iowa} \\
  Iowa City, IA, USA \\
  weiran-wang@uiowa.edu
}
}

\maketitle

\begin{abstract}
Large audio-language models (LALMs) frequently hallucinate by overriding acoustic evidence with language priors. While contrastive decoding (CD) offers training-free mitigation, existing methods rely on blunt perturbations like masking or noise, leaving structured audio transformations unexplored. We explore this design space by evaluating a diverse library of targeted audio perturbations and adaptively selecting the optimal negative branch for each task and example. First, we improve upon earlier prompt engineering by showing that a simple binary yes/no constraint reduces the model's tendency to falsely confirm absent audio features. Second, evaluating our library across temporal, spectral, frequency, and amplitude domains reveals that optimal transformations are highly task-dependent; for instance, reversing the audio array disrupts temporal coherence, raising accuracy on the temporal order task from $74.7\%$ to $81.4\%$. Finally, we trained a light-weight perturbation selector on model hidden states to dynamically route negative branches, yielding an additional $+4.3\%$ gain on the existence task.\footnote{Code available at \url{https://github.com/aarongrace/adaptive-lalm-cd}.}% chktex 13
\end{abstract}


\section{Introduction}

Large audio-language models (LALMs)~\cite{tang2024salmonn,deshmukh2023pengi,gong2024ltu,yu2025salmonnomni,lu2025desta25audio,hu2024wavllm} score well on benchmarks but routinely hallucinate~\cite{kuan2025can,kuan2024usmq,zhao2026halluaudio,cheng2025ahabench,sungbin2025avhbenchcrossmodalhallucinationbenchmark}. These errors typically emerge during decoding when strong textual priors dominate the output space, causing the model to generate linguistically probable text that overrides the actual acoustic reality.
Contrastive decoding (CD) offers a training-free remedy~\cite{li2022contrastive}. By amplifying the log-probability difference between an expert model (normal input branch) and a weaker, perturbed negative branch, CD penalizes tokens driven purely by language priors. However, current applications rely on a limited set of negative branches---such as total audio masking or basic additive noise~\cite{hsu2025reducing}---leaving the broader design space of audio perturbations largely unexplored.

In this work, we perform a systematic exploration of this uncharted design space by evaluating a diverse library of targeted audio perturbations for CD. Since diverse audio tasks depend on varying acoustic characteristics, the design of a negative branch must be task-specific. Contrastive examples must remain meaningful without introducing severe acoustic distortions that may induce further hallucinations.% chktex 13

We evaluate our method across two LALMs: Qwen2-Audio-7B-Instruct~\cite{chu2024qwen2} and Audio Flamingo 3~\cite{goel2025af3} (henceforth AF3). Testing spans four tasks: Clotho-AQA~\cite{lipping2022clotho}, along with the existence, temporal order, and object attribute variants of the Audio Hallucination dataset~\cite{kuan2025can} (henceforth AH Existence, AH Order, and AH Attribute). % chktex 8
Our contributions are:
\begin{itemize}
  \item \textbf{Prompt calibration.} Constraining model outputs to a single yes/no token calibrates the model's innate affirmative bias, raising AH Existence accuracy by $+11$\% before any CD---more than four times the prompt engineering gain of prior work on the same model.
  \item \textbf{Perturbation library.} We introduce an extensive perturbation library for audio CD, comprising 105 perturbations across 38 types covering temporal, spectral, frequency, and amplitude transformations. We show that optimal perturbations are strongly task-dependent: for AF3, a reverse-audio branch achieves $+6.7$\% on AH Order (74.7\%$\to$81.4\%) and a pitch-shift branch achieves $+4.4$\% on AH Existence (69.5\%$\to$73.9\%).  
  \item \textbf{Adaptive selector.} Because the optimal perturbation depends on both the task and the individual example, we train a lightweight selector on model hidden states to route each input to its best negative branch, adding $+4.3$\% over the best fixed branch on AH Existence with Qwen2, with $+9.7$\% of oracle headroom remaining.
\end{itemize}

\begin{figure*}[tp!]
\centering
\includegraphics[width=\textwidth]{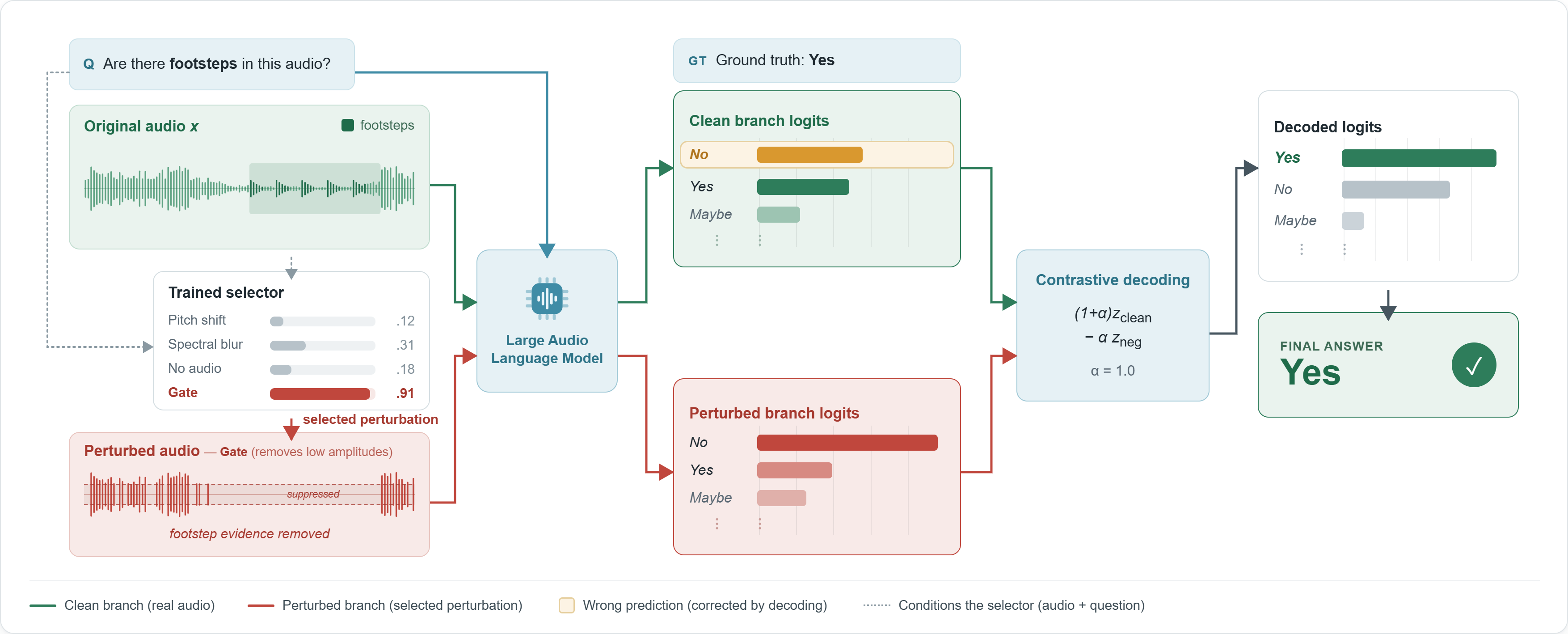}
\caption{System overview. The selector chooses a perturbation based on text and audio embeddings. Both branches are forwarded through the LALM, and contrastive correction (Eq.\,\eqref{eq:cd}) is applied at decoding.}% chktex 13
\label{fig:system}
\end{figure*}

\section{Related Work}

\textbf{Text and vision contrastive decoding.} Li et al.~\cite{li2022contrastive} introduced CD for text generation, contrasting a large expert LM against a smaller amateur LM to suppress repetition and topic drift. VCD~\cite{leng2024vcd} adapts this to vision-language models, using noise-distorted visual inputs to construct the negative branch and mitigate object hallucination. While DoLa~\cite{chuang2024dola} contrasts representations across layers internally, AVCD~\cite{jung2025avcd} extends CD to the trimodal audio-visual setting. Most relevant to our work, VACoDe~\cite{kim2024vacode} dynamically selects the most contrastive visual augmentation per query via softmax distance; we evaluate an analogous criterion in Section~\ref{sec:distance}.

\textbf{Contrastive decoding with audio language models.} Our work builds %directly 
on the foundational Audio-Aware Decoding (AAD) framework~\cite{hsu2025reducing}, which pioneered inference-time CD for LALMs. AAD elegantly isolates persistent language priors by contrasting standard predictions against a negative branch with no-audio; pairing this token-level subtraction with an attention-directing prompt yields substantial, training-free gains across SALMONN and Qwen2-Audio on Clotho-AQA and AH Existence. Temporal Contrastive Decoding~\cite{li2026temporalcontrastivedecodingtrainingfree} extends this approach by replacing the no-audio branch with a temporally blurred view and applying an uncertainty-gated logit update, producing improvements on MMAU and AIR-Bench. Lin et al.~\cite{lin2025contrastive} further characterized these gains via a transition matrix, showing CD reliably corrects false audio-absence claims but cannot fix highly confident incorrect beliefs. We extend this foundation to a structured library of task-oriented perturbations, with a lightweight selector that routes among them per example.

\section{Method}
Fig.~\ref{fig:system} illustrates our overall architecture.

\subsection{Problem Setup and Contrastive Decoding}

Let $x$ be an audio clip, $q$ a question prompt, and $y$ an answer. The LALM produces next-token logits
\begin{equation}
  z_t = f_\theta(y_{<t},\, q,\, x),
\end{equation}
where $f_\theta$ denotes the model. A perturbation $s$ defines $T_s(x)$. The contrastive branch logits
\begin{equation}
  z^-_{t,s} = f_\theta(y_{<t},\, q,\, T_s(x))
\end{equation}
are subtracted at each decoding step via a logits processor:
\begin{equation}
  \tilde{z}_{t,s} = (1 + \alpha)\,z_t - \alpha\,z^-_{t,s},
  \label{eq:cd}
\end{equation}
where $\alpha$ controls subtraction strength.

\subsection{Perturbation Library}

We construct a library which covers 105 perturbations across 38 types in six families. Frequency filtering uses fifth-order Butterworth filters (SciPy~\cite{Virtanen_2020}); pitch shifting, time stretching, spectral operations, and source separation use librosa~\cite{mcfee2015}.

\subsubsection{Temporal}
\begin{itemize}
  \item \textbf{Reverse:} Reverses the time axis of the waveform.
  \item \textbf{Time Stretch:} Phase-vocoder stretch at $0.4\times$ and $2.5\times$.
  \item \textbf{Segment Shuffle:} Waveform divided into $K \in \{10,50,200\}$ blocks, randomly reordered.
  \item \textbf{Segment Reverse:} Each of $K \in \{10,50\}$ blocks reversed in-place.
  \item \textbf{Dropout:} Samples zero-masked at $p \in \{0.4,0.7\}$.
  \item \textbf{Time Mask:} Contiguous silence intervals up to 15\% of audio duration.
  \item \textbf{Repeat Segment:} A 20\% slice (start or middle) looped to fill the original duration.
\end{itemize}

\subsubsection{Frequency Filters}
\begin{itemize}
  \item \textbf{Low-Pass / High-Pass:} Hard roll-offs at 250--1000\,Hz and 1--6\,kHz respectively.
  \item \textbf{Bandpass / Bandstop:} Isolates (bandpass, ten bands) or notches (bandstop, three bands) sub-bands spanning 50\,Hz--8\,kHz; e.g.\ bass (50--300\,Hz), mid (500--2000\,Hz), treble (3--8\,kHz).
  \item \textbf{Frequency Mask:} Multiple random STFT bands zeroed and reconstructed via ISTFT\@.
\end{itemize}

\subsubsection{Spectral}
\begin{itemize}
  \item \textbf{Pitch Shift:} $\pm4$ to $\pm24$ semitones; duration preserved.
  \item \textbf{Spectral Noise:} Gaussian noise injected into STFT magnitude ($\sigma_{\mathrm{spec}} \in \{0.1,0.3,0.6,1.0\}$ relative to magnitude RMS); phase retained.
  \item \textbf{Spectral Blur:} Sequential 1D Gaussian smoothing of the STFT magnitude along time and frequency. The standard deviation $\sigma_{\mathrm{bin}}$, measured in STFT bins, controls the blur width and is shared across both axes ($\sigma_{\mathrm{bin}} \in \{5,15,25\}$).
  \item \textbf{Spectral Reverse / Segment Reverse / Segment Shuffle:} Direct analogues of the temporal operations applied to STFT frame columns and reconstructed via ISTFT\@.
  \item \textbf{Harmonic Removal / Percussive Removal:} Librosa median-filter separation on the spectrogram (margin 3.0); each retains only the tonal or transient component.
\end{itemize}

\subsubsection{Amplitude \& Dynamics}
\begin{itemize}
  \item \textbf{Hard Clip:} Truncates amplitude to $\pm\mathrm{thr}$ ($\mathrm{thr} \in \{0.1,0.2\}$).
  \item \textbf{Quantize:} Bit depth reduced to 2--4 bits.
  \item \textbf{Compress:} 10:1 or 20:1 ratio applied above threshold.
  \item \textbf{Gate / Gate Inverted:} Zeros samples below (gate, $\mathrm{thr} \in \{0.30,0.50,0.65,0.75\}$) or above (gate inverted, $\mathrm{thr} \in \{0.10,0.20,0.30,0.50\}$) a linear amplitude threshold.
  \item \textbf{Gate Soft / Gate Inverted Soft:} Same as above with $-12$ to $-45$\,dB attenuation instead of hard zeroing.
  \item \textbf{Normalize Chunks:} Peak-normalizes $K \in \{10,50\}$ independent temporal chunks.
  \item \textbf{Resample Low / Bit Crush:} Downsample to 2--8\,kHz then upsample; Bit Crush additionally reduces bit depth.
\end{itemize}

\subsubsection{Environmental}
\begin{itemize}
  \item \textbf{Reverb / Echo:} Multi-tap comb delays at 50--300\,ms with configurable decay.
  \item \textbf{Phone Filter:} Bandpass restricted to 300--3400\,Hz.
  \item \textbf{Underwater:} Low-pass at 400\,Hz followed by reverb.
\end{itemize}

\subsubsection{Additive Noise}
\begin{itemize}
  \item \textbf{White Noise / Colored Noise:} Gaussian or pink/brown noise at $\sigma_{\mathrm{wave}} \in \{0.3\text{--}1.0\}$ relative to waveform amplitude.
\end{itemize}

\subsection{Oracle Labels and Selector Training}

\begin{figure}[t]
\centering
\begin{tikzpicture}[
  font=\small,
  >=Stealth,
  cell/.style={draw, minimum width=1.18cm, minimum height=0.42cm,
               align=center, inner sep=1pt, font=\footnotesize},
  rbox/.style={draw, rounded corners=2pt, inner sep=4pt, align=center, font=\small},
]

% ── Input ────────────────────────────────────────────────────
\node[rbox] (inp) at (0,0) {audio + question~$i$};
\draw[->] (inp.south) -- (0,-0.80)
  node[midway, right=2pt, font=\scriptsize\itshape]
  {offline CD eval.\ per candidate};

% ── Perturbation name row ────────────────────────────────────
\node[cell] at (-2.36,-1.05) {no-audio};
\node[cell] at (-1.18,-1.05) {pitch};
\node[cell] at ( 0.00,-1.05) {blur};
\node[cell] at ( 1.18,-1.05) {gate};
\node[font=\footnotesize] at (2.36,-1.05) {$\cdots$};

% ── Result row ───────────────────────────────────────────────
\node[cell, fill=green!15] at (-2.36,-1.55) {$\checkmark$};
\node[cell, fill=red!10]   at (-1.18,-1.55) {$\times$};
\node[cell, fill=green!15] at ( 0.00,-1.55) {$\checkmark$};
\node[cell, fill=red!10]   at ( 1.18,-1.55) {$\times$};
\node[font=\footnotesize]  at ( 2.36,-1.55) {$\cdots$};

% ── Oracle target ─────────────────────────────────────────────
\draw[->] (0,-1.79) -- (0,-2.25);
\node[font=\footnotesize] at (0,-2.45)
  {oracle target:\enspace $M_i = [\,1,\;0,\;1,\;0,\;\ldots\,]$};

% Dashed separator
\draw[dashed, gray!70] (-3.1,-2.75) -- (3.1,-2.75);

% ── Cached hidden state → selector → scores ──────────────────
\node[rbox, font=\footnotesize] (hi)  at (-2.4,-3.25) {cached~$h_i$};
\node[rbox, minimum width=1.8cm, font=\footnotesize] (mlp) at (0,-3.25)
  {selector MLP};
\node[rbox, font=\scriptsize] (sc) at (2.65,-3.25)
  {$[\,0.82,\;0.21,\;0.74,\;\ldots\,]$};

\draw[->] (hi)  -- (mlp);
\draw[->] (mlp) -- (sc);

\node[font=\scriptsize, text=gray!80!black] at (0,-3.72)
  {train: BCE \quad|\quad infer: argmax};

\end{tikzpicture}
\vspace{-2ex}
\caption{Selector training. Offline CD evaluation yields a multi-hot correctness vector $M_i$ per example; the MLP is trained with BCE on cached hidden states and selects the negative branch by argmax at inference.}%
\label{fig:training}
\end{figure}
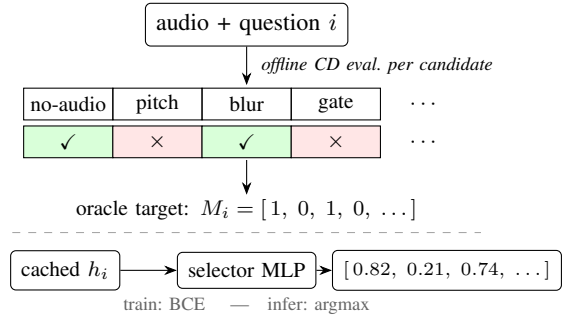

For each example $i$ in the evaluation set $\mathcal{D}$ and perturbation $s$, let $M_{i,s} = \mathbf{1}[\hat{y}_{i,s} = y_i]$ indicate correctness. The oracle accuracy establishes an upper bound assuming an ideal per-example selector:
\begin{equation}
  A_{\mathrm{oracle}} = \frac{1}{|\mathcal{D}|}\sum_{i \in \mathcal{D}} \max_s M_{i,s}
\end{equation}
We train the selector (Fig.~\ref{fig:training}) using the vector $M_i$ as a multi-label binary target, treating all correct perturbations as positives. The selector itself is a lightweight neural network trained directly on the LALM's internal hidden states to predict these targets. At inference, the perturbation with the highest predicted score becomes the negative branch used for contrastive decoding.

\section{Experimental Setup}

\textbf{Datasets.} Table~\ref{tab:datasets} summarizes the four evaluation settings. The AH benchmark~\cite{kuan2025can} provides three yes/no grounding tasks. AH Existence uses 10{,}800 questions, created from 600 combinations of one background and three foreground events, with multiple present/absent questions per combination. AH Order and AH Attribute use compositional scenes from CompA~\cite{ghosh2024compa}, testing temporal ordering and sound-attribute reasoning respectively. Lacking standard splits, we construct five 70/15/15 splits per task and average results across them. We partition at the audio-composition level, keeping all questions from the same composition together. For AH Order and AH Attribute, no audio files are shared between training and held-out partitions. AH Existence reuses 1{,}120 source clips across its 600 combinations, making strict clip-level separation impossible. We therefore keep combinations disjoint and minimize train/held-out clip overlap, yielding Jaccard similarities of 0.12--0.13.
Clotho-AQA~\cite{lipping2022clotho} provides crowd-sourced yes/no questions over FreeSound clips~\cite{Font2013freesound}, predominantly covering existence and attribute tasks.

\begin{table}[t]
\caption{Dataset summary (yes/no subsets only).}%
\label{tab:datasets}
\centering
\vspace{-1ex}
\begin{tabular}{llrl}
\toprule
Dataset & Task & Examples & Source \\
\midrule
AH Existence  & Existence Y/N & 10{,}800 & Synthetic composite~\cite{kuan2025can} \\
AH Order      & Order Y/N     & 3{,}078  & CompA~\cite{ghosh2024compa} \\
AH Attribute  & Attribute Y/N & 1{,}599  & CompA~\cite{ghosh2024compa} \\
Clotho-AQA    & Mixed Y/N     & 7{,}959  & FreeSound~\cite{Font2013freesound} \\
\bottomrule
\end{tabular}
\end{table}

\textbf{Models.} We evaluate on two LALMs: Qwen2-Audio-7B-Instruct~\cite{chu2024qwen2} (Whisper-large-v2~\cite{radford2023whisper} encoder, Qwen-7B backbone) and AF3~\cite{goel2025af3} (AF-Whisper encoder, Qwen2.5-7B backbone). % chktex 8

\textbf{Metrics.} The primary metric is accuracy: a prediction is correct if the argmax token matches the ground-truth yes/no label. All main perturbation and selector experiments use the constrained prompt (Section~\ref{sec:prompts}).

\section{Results and Analysis}\label{sec:results}

\subsection{Prompt Engineering and Calibration}\label{sec:prompts}

Table~\ref{tab:prompts} (a) details our output prompts. Prior work~\cite{hsu2025reducing} added an attention-directing prefix; we extend this by appending an explicit yes/no constraint---the \textbf{constrained prompt} used in all main experiments---significantly boosting performance.

\begin{table}[t]
\caption{Prompt engineering comparison. \textbf{(a)}~Prompts evaluated. \textbf{(b)}~Accuracy (\%) under original and AAD decoding ($\alpha$=1.0).}%
\label{tab:prompts}\label{tab:prompt}
\centering
\vspace{-1ex}
\noindent\textbf{(a)}\\[3pt]
\begin{tabular}{p{1.8cm}p{5.7cm}}
\toprule
Prompt & Text \\
\midrule
AAD Prompt~\cite{hsu2025reducing} & ``Focus on the given audio and answer the following question.''\ % chktex 38
\\
Our Prompt & ``Focus on the given audio and answer the following question \textbf{with exactly one word: yes or no.}''\ % chktex 38
\\
\bottomrule
\end{tabular}

\vspace{6pt}
\noindent\textbf{(b)}\\[3pt]
\begin{tabular}{@{}ll@{\hspace{6pt}}rrrr@{}}
\toprule
 & & \multicolumn{2}{c}{No Perturbation} & \multicolumn{2}{c}{AAD (No-Audio)} \\
\cmidrule(lr){3-4}\cmidrule(lr){5-6}
Model & Dataset & \shortstack{AAD\\Prompt} & \shortstack{Our\\Prompt} & \shortstack{AAD\\Prompt} & \shortstack{Our\\Prompt} \\
\midrule
Qwen2 & AH Existence & 56.9 & 67.9 & 70.9 & \textbf{72.4} \\
Qwen2 & AH Order     & 50.4 & 51.2 & \textbf{53.2} & 52.8 \\
Qwen2 & AH Attribute & 49.9 & \textbf{51.0} & 50.7 & \textbf{51.0} \\
Qwen2 & Clotho-AQA   & 72.5 & 76.1 & 78.6 & \textbf{79.6} \\
\midrule
AF3   & AH Existence & 66.9 & 69.5 & 72.2 & \textbf{73.1} \\
AF3   & AH Order     & 77.6 & 74.7 & \textbf{79.5} & 76.7 \\
AF3   & AH Attribute & 55.7 & \textbf{56.0} & 55.7 & 55.5 \\
AF3   & Clotho-AQA   & 81.7 & \textbf{82.6} & 81.8 & 82.1 \\
\bottomrule
\end{tabular}
\end{table}

Under the standard AAD prompt, Qwen2 assigns ``yes'' to 90.4\% of AH Existence examples. On this balanced dataset, this yields a $+40.4$\% affirmative bias (predicted-yes rate minus 50\%). The constrained prompt reduces this bias to $+21.0$\%, raising accuracy from 56.9\% to 67.9\% ($+11.0$\%). Applying no-audio AAD ($\alpha=1.0$) further shrinks residual bias to $+1.8$\%, reaching 72.4\% accuracy. The same additive pattern holds on Clotho-AQA ($72.5\% \to 76.1\%$ from the prompt; $76.1\% \to 79.6\%$ from no-audio contrastive decoding).

Accuracy across both models and all tasks is in Table~\ref{tab:prompt}~(b). All subsequent experiments use the constrained prompt.

\subsection{Alpha Sensitivity}\label{sec:alpha}

\begin{figure}[t]
\centering
\includegraphics[width=0.88\columnwidth]{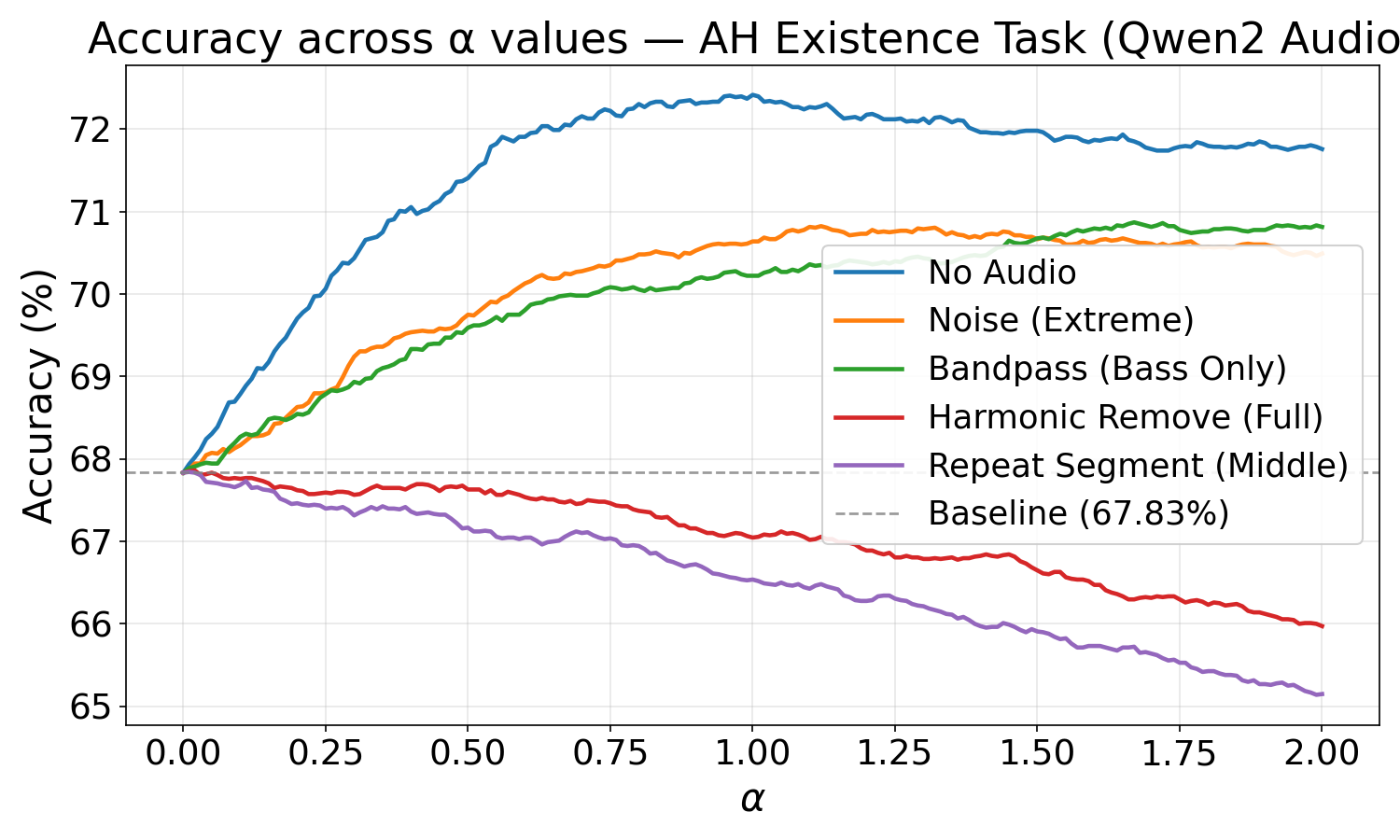}
\vspace{-6pt}
\caption{Accuracy across $\alpha$ values. Helpful perturbations improve monotonically up to $\alpha\approx 1.0$, sometimes with marginal gains just beyond; harmful perturbations degrade monotonically throughout.}% chktex 13
\label{fig:alpha}
\end{figure}

Sweeping $\alpha \in [0, 2]$ across five representative perturbations (Fig.~\ref{fig:alpha}) reveals two behaviors. Helpful perturbations---no-audio, noise ($\sigma_{\mathrm{wave}}{=}0.6$), and bandpass (50--300\,Hz)---improve with $\alpha$, peaking near or slightly above $\alpha=1.0$ with marginal gains before plateauing. Harmful perturbations---harmonic remove (full) and repeat segment (middle)---degrade monotonically from the outset. We avoid $\alpha>1.0$: beyond this point the correction term outweighs the expert branch, making gains harder to interpret. We fix $\alpha=1.0$ for remaining experiments.

\subsection{Individual Perturbation Rankings}\label{sec:fixed}
\begin{table*}[t]
\caption{Perturbation accuracy (\%) at $\alpha=1.0$ (constrained prompt). Top four and bottom two perturbations shown per task; clean baseline listed separately. AH Attribute omitted; results are reported in text. Clotho-AQA results pooled across train, validation, and test splits. }%
\label{tab:perturbations}
\centering
\begingroup
\setlength{\tabcolsep}{2pt}
\begin{tabularx}{\textwidth}{@{}p{0.5cm}@{\hspace{8pt}} Yr | Yr | Yr @{\hspace{3pt}}||@{\hspace{3pt}} Yr | Yr | Yr@{}}% chktex 44
\toprule
 & \multicolumn{6}{c}{Qwen2-Audio-7B-Instruct} & \multicolumn{6}{c}{AF3} \\% chktex 8
\cmidrule(lr){2-7}\cmidrule(lr){8-13}
 & \multicolumn{2}{c|}{AH Existence} & \multicolumn{2}{c|}{AH Order} & \multicolumn{2}{c}{Clotho-AQA} & \multicolumn{2}{c|}{AH Existence} & \multicolumn{2}{c|}{AH Order} & \multicolumn{2}{c}{Clotho-AQA} \\
\midrule
\multicolumn{2}{@{}l}{Original} & 67.8 & & 51.2 & & 76.9 & & 69.5 & & 74.7 & & \textbf{86.7} \\
\multicolumn{2}{@{}l}{Oracle}   & 86.4 & & 61.6 & & 90.1 & & 85.0 & & 90.6 & & 93.8 \\
\midrule
\#1  & No-Audio & 72.4 & Reverse & 53.5 & No-Audio & 79.4 & Pitch shift & 73.9 & Reverse & 81.4 & Noise & 86.7 \\
     & \multicolumn{2}{l|}{} & \multicolumn{2}{l|}{} & \multicolumn{2}{l}{} & \multicolumn{2}{l|}{{\small(up two octaves)}} & \multicolumn{2}{l|}{} & \multicolumn{2}{l}{{\small($\sigma_{\mathrm{wave}}{=}1.0$)}} \\[2pt]
\#2  & Noise & 70.6 & Segment shuffle & 52.9 & Gate & 78.1 & Gate & 73.8 & Spectral reverse & 80.0 & Bandstop & 86.7 \\
     & \multicolumn{2}{l|}{{\small($\sigma_{\mathrm{wave}}{=}0.6$)}} & \multicolumn{2}{l|}{{\small(10 seg)}} & \multicolumn{2}{l}{{\small(thr=0.75)}} & \multicolumn{2}{l|}{{\small(thr=0.75)}} & \multicolumn{2}{l|}{} & \multicolumn{2}{l}{{\small(500\,Hz--2\,kHz)}} \\[2pt]
\#3  & Noise & 70.5 & No-Audio & 52.8 & Bandpass & 77.9 & Segment shuffle & 73.8 & Segment shuffle & 79.1 & Low pass & 86.7 \\
     & \multicolumn{2}{l|}{{\small($\sigma_{\mathrm{wave}}{=}0.5$)}} & \multicolumn{2}{l|}{} & \multicolumn{2}{l}{{\small(2--3.5\,kHz)}} & \multicolumn{2}{l|}{{\small(200 seg)}} & \multicolumn{2}{l|}{{\small(10 seg)}} & \multicolumn{2}{l}{{\small(1\,kHz)}} \\[2pt]
\#4  & Pitch shift & 70.3 & Time stretch & 52.7 & Gate & 77.9 & Gate inv. & 73.7 & Spec.\ seg.\ shuffle & 78.9 & Gate & 86.6 \\
     & \multicolumn{2}{l|}{{\small(down octave)}} & \multicolumn{2}{l|}{{\small(2.5$\times$)}} & \multicolumn{2}{l}{{\small(thr=0.65)}} & \multicolumn{2}{l|}{{\small(thr=0.10)}} & \multicolumn{2}{l|}{{\small(10 seg)}} & \multicolumn{2}{l}{{\small(thr=0.65)}} \\[2pt]
$\cdots$ & \multicolumn{2}{c}{$\cdots$} & \multicolumn{2}{c}{$\cdots$} & \multicolumn{2}{c}{$\cdots$} & \multicolumn{2}{c}{$\cdots$} & \multicolumn{2}{c}{$\cdots$} & \multicolumn{2}{c}{$\cdots$} \\
\#104 & Segment shuffle & 66.6 & Clip & 51.0 & Spectral blur & 75.7 & Spectral noise & 68.9 & Resample low & 72.7 & Dropout & 85.7 \\
      & \multicolumn{2}{l|}{{\small(200 seg)}} & \multicolumn{2}{l|}{{\small(thr=0.2)}} & \multicolumn{2}{l}{{\small($\sigma_{\mathrm{bin}}{=}25$)}} & \multicolumn{2}{l|}{{\small($\sigma_{\mathrm{spec}}{=}0.1$)}} & \multicolumn{2}{l|}{{\small(8\,kHz)}} & \multicolumn{2}{l}{{\small($p{=}0.4$)}} \\[2pt]
\#105 & Repeat segment & 66.5 & Noise & 50.8 & Spectral blur & 75.5 & Quantize & 68.6 & Normalize chunks & 71.8 & Reverb & 85.5 \\
      & \multicolumn{2}{l|}{{\small(middle)}} & \multicolumn{2}{l|}{{\small($\sigma_{\mathrm{wave}}{=}1.0$)}} & \multicolumn{2}{l}{{\small($\sigma_{\mathrm{bin}}{=}15$)}} & \multicolumn{2}{l|}{{\small(4-bit)}} & \multicolumn{2}{l|}{{\small(10 chunks)}} & \multicolumn{2}{l}{{\small(dec=0.95, 100\,ms)}} \\[2pt]
\bottomrule
\end{tabularx}
\endgroup
\end{table*}

Per-setting rankings at $\alpha=1.0$ (Table~\ref{tab:perturbations}) reveal task-specific patterns.

\textbf{AH Existence.} For Qwen2, no-audio is the dominant branch at 72.4\%, which is 4.6\% above the original baseline without perturbation (67.8\%). AF3 is not dominated by no-audio: pitch shift (up two octaves) leads at 73.9\%, 4.4\% above the original baseline (69.5\%) and above no-audio at 73.1\%. Conversely, Qwen2's worst branches---segment shuffle (200 segments, 66.6\%) and repeat segment (middle, 66.5\%)---fall below the unmodified baseline (67.8\%). Because neither produces a reliable absence contrast, they provide no useful contrastive signal for existence.

\textbf{AH Order.} This pattern reverses for AH Order, where temporal shuffling provides a useful contrastive signal. Full waveform reversal leads for both models: $+2.4$\% for Qwen2 (51.2\%$\to$53.5\%) and $+6.7$\% for AF3 (74.7\%$\to$81.4\%). Temporal inversion forms an ideal negative branch for order-sensitive questions.

\textbf{AH Attribute.} Neither model improves meaningfully under contrastive decoding. Qwen2 rises from 51.0\% to 53.0\% with $0.4\times$ time stretch (52.9\% at $2.5\times$), while spectral noise performs worst at 50.5\% ($\sigma_{\mathrm{spec}}=0.6$) and 50.2\% ($\sigma_{\mathrm{spec}}=1.0$). AF3 rises only from 56.0\% to 56.5\% with either $0.4\times$ time stretch or 10:1 compression, while high-pass filtering at 6\,kHz and gating at threshold 0.50 reduce accuracy to 52.0\% and 51.8\%, respectively. We therefore omit AH Attribute from Table~\ref{tab:perturbations}.

\textbf{Clotho-AQA.} AF3 is already at ceiling (86.7\%) and no contrastive branch improves on the original. For Qwen2, the no-audio branch leads but gains are modest.

\subsection{Distance-Based Perturbation Selection}\label{sec:distance}

\begin{figure}[t]
  \centering
  \includegraphics[width=\columnwidth]{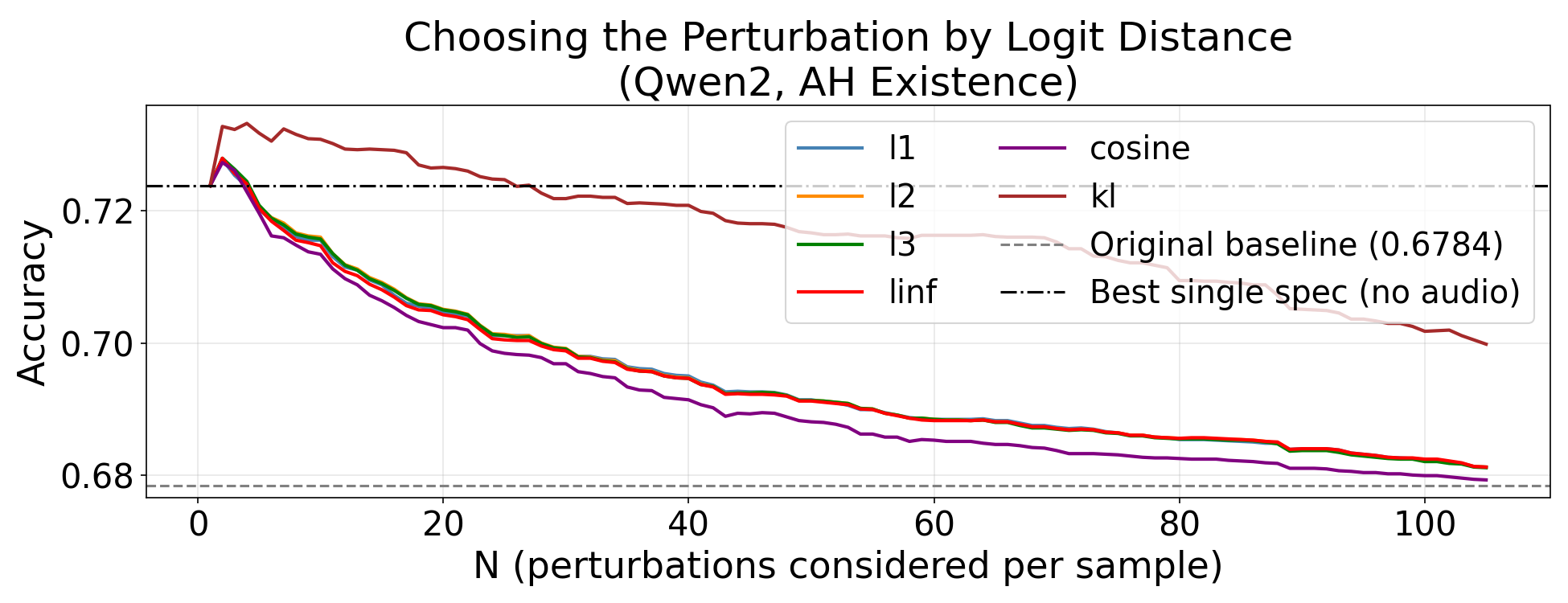}
  \caption{Distance-based branch selection on Qwen2 AH Existence ($\alpha=1.0$). Each line shows accuracy for a given softmax distance metric as $N$ ranges from 1 to 105. Reference line: fixed no-audio baseline (72.4\%). }%
\label{fig:distance}
\end{figure}

VACoDe~\cite{kim2024vacode} selects the negative branch that maximizes softmax divergence from the original prediction. In the visual domain, VACoDe found $L_2$ to be the strongest metric. We evaluated the analogous strategy for audio across six distance metrics ($L_1$, $L_2$, $L_3$, $L_\infty$, cosine, KL); among our candidates, KL divergence yielded a small benefit. Applying distance-based selection across the full perturbation pool yields unstable results; we therefore report performance restricted to the top-$N$ perturbations by aggregate accuracy (Fig.~\ref{fig:distance}), the strongest variant we found. Sampling from this top-performing pool and restricting to weaker perturbations at $\alpha=0.5$ both performed strictly worse. Distance-based selection consistently favors the most extreme noise and pitch-shift variants, suggesting that logit divergence rewards severe acoustic distortion rather than contrastive utility.

\subsection{Adaptive Selector}\label{sec:selector}

\begin{table}[t]
\caption{Per-split selector test accuracy (\%) on AH Existence using optimal configuration: $N=4$, 3-layer MLP [512, 256, 128], concatenated last-token hidden states from first, middle, and final LALM layers, label smoothing $\epsilon=0.25$, feature noise $=0.10$, input dropout $=0.05$, and $\alpha=1.0$.}
\label{tab:split_variance}
\centering
\begin{tabular}{lrrrrrrr}
\toprule
Model & S1 & S2 & S3 & S4 & S5 & Mean & Std. \\
\midrule
Qwen2 & 76.11 & 77.22 & 77.41 & 76.65 & 76.05 & \textbf{76.69} & 0.62 \\
AF3   & 76.67 & 76.61 & 76.30 & 75.93 & 76.36 & \textbf{76.37} & 0.29 \\
\bottomrule
\end{tabular}
\end{table}
The adaptive selector routes each audio-question pair to one of $N$ candidate negative branches. It operates on hidden states from the expert forward pass and requires no additional LALM computation. Offline label generation requires evaluating each candidate perturbation, but at inference, the selector adds only a lightweight MLP over the two LALM forward passes already required by fixed-branch CD. We evaluate four design choices using Qwen2 on AH Existence ($\alpha=1.0$): candidate pool size, head architecture, input features, and regularization. Each ablation holds the remaining settings fixed at their best preceding configuration. We report final selector performance across five splits in Table~\ref{tab:split_variance}, with test accuracy of $76.69\pm0.62$\% for Qwen2 and $76.37\pm0.29$\% for AF3.

\subsubsection{Number of Candidate Perturbations ($N$)}

Expanding the candidate pool continuously raises oracle performance, reaching 86.4\% for Qwen2 and 85.0\% for AF3 at $N=60$ (Table~\ref{tab:nspec}). Conversely, selector accuracy peaks early at $N=4$ for both models (76.7\% and 76.4\%) and then declines. This divergence widens the oracle-selector gap to 11.1\% at $N=60$ because larger pools dilute the training signal across the ${\approx}7{,}500$ available samples. A greedy search over all 105 perturbations shows steep diminishing returns. Maximum coverage requires $N\approx51$, but performance largely plateaus by $N=10$.

\begin{table}[t]
\caption{Oracle and best selector accuracy (\%) by number of candidate perturbations $N$. Here Gap $=$ Oracle $-$ Selector.}%
\label{tab:nspec}
\centering
\begin{tabular}{r rr r@{\hspace{10pt}} rr r}
\toprule
 & \multicolumn{3}{c}{Qwen2} & \multicolumn{3}{c}{AF3} \\
\cmidrule(lr){2-4}\cmidrule(lr){5-7}
$N$ & Oracle & Selector & Gap & Oracle & Selector & Gap \\
\midrule
0  & 67.8 & ---           & --- & 69.5 & ---           & --- \\
1  & 72.4 & 72.4          & 0.0 & 73.9 & 73.9          & 0.0 \\
3  & 82.9 & 76.4          & 6.5 & 80.4 & 76.1          & 4.3 \\
4  & 83.5 & \textbf{76.7} & 6.8 & 81.3 & \textbf{76.4} & 4.9 \\
6  & 84.4 & 76.5          & 7.9 & 82.2 & 76.4          & 5.8 \\
10 & 85.2 & 76.1          & 9.1 & 83.2 & 76.2          & 7.0 \\
20 & 85.9 & 75.8          &10.1 & 84.2 & 76.2          & 8.0 \\
30 & 86.1 & 75.7          &10.4 & 84.6 & 76.0          & 8.6 \\
60 & 86.4 & 75.3          &11.1 & 85.0 & 75.9          & 9.1 \\
\midrule
\multicolumn{7}{@{}l}{\small\textit{Selected candidate perturbations ($N=4$, $\alpha=1.0$):}} \\
\multicolumn{7}{@{}p{8.2cm}}{\small \textit{Qwen2}: No-audio; Noise ($\sigma_{\mathrm{wave}}{=}1.0$); High pass (6\,kHz); Spectral blur ($\sigma_{\mathrm{bin}}{=}15$)} \\
\multicolumn{7}{@{}p{8.2cm}}{\small \textit{AF3}: Gate inverted (thr$=0.10$); Gate (thr$=0.75$); Original; Pitch shift ($+24$ semitones)} \\
\bottomrule
\end{tabular}
\end{table}

\subsubsection{Head Architecture}
We swept MLP depth and width; a 3-layer head with hidden dimensions \mbox{[512, 256, 128]} proved optimal at \textbf{76.7\%}, with performance degrading consistently beyond 3 layers due to overfitting.

\subsubsection{Input Features}\label{sec:selector_ah}

Table~\ref{tab:ablation} ablates input feature configurations on Qwen2 AH Existence ($N=4$, $\alpha=1.0$, 5 balanced splits) using 3-layer MLP selector heads.

\textbf{The last-step hidden representation is the critical feature.} Mean-pooled LLM hidden states plateau at 72.3--72.7\%, near the no-audio baseline of 72.4\%, while \textbf{last-token, last-layer} reaches 76.3\%. In a causal decoder, the final non-padding token attends to the full input, including the system prompt, audio tokens, and question. Mean pooling dilutes this information by averaging over earlier positions with only partial context.

\textbf{Multi-layer extraction improves further.} Concatenating last-token states from first, middle, and final layers was the best approach, reaching \textbf{76.7\%} ($+4.3$\% over fixed no-audio, $+8.9$\% over the unmodified model); earlier layers contribute complementary information the final layer discards.

\textbf{Separate audio features are redundant.} Text-token LLM hidden states alone provide the strongest routing signal. Raw or projected audio embeddings match mean-pooled text-token states (72.5--72.6\%), and appending separate audio features to any text-token representation yields no reliable gain, and cross-attention (last-token $\oplus$ audio) reaches 76.4\%, essentially matching the last-token text representation alone. External audio re-injection likely adds little because text-token hidden states already encode substantial acoustic information through multimodal attention. This generalizes to AF3, where the best selector reaches 76.4\% ($N=4$), versus the 73.1\% no-audio baseline and 69.5\% original.

\begin{table}[t]
\caption{Input features ablation for the selector (Qwen2, AH Existence, $N=4$, $\alpha=1.0$, avg.\ over 5 balanced splits).}%
\label{tab:ablation}
\centering
\begin{tabular}{@{}p{5.2cm}@{\hspace{4pt}}r@{}}
\toprule
\multicolumn{2}{@{}l}{\textit{Baselines} \textbar{} Original: 67.8 \enspace Best perturbation: 72.4 \enspace Oracle: 86.4} \\
\midrule
Feature configuration & Acc (\%) ($\Delta$ vs.\ no-audio) \\
\midrule
\multicolumn{2}{c}{\textit{LLM hidden states only}} \\
Mean pool, last layer & 72.3 ($-$0.1) \\
Mean pool, all layers & 72.7 (+0.3) \\
Last token, last layer & 76.3 (+3.9) \\
Last token $\oplus$ all-layer mean & 76.5 (+4.1) \\
\textbf{Last token, first/mid/last layers concat.} & \textbf{76.7 (+4.3)} \\
\midrule
\multicolumn{2}{c}{\textit{Audio encoder features only}} \\
Projected audio mean & 72.5 (+0.1) \\
Raw audio mean & 72.6 (+0.2) \\
\midrule
\multicolumn{2}{c}{\textit{LLM hidden states $\oplus$ audio (concat)}} \\
Last token $\oplus$ projected audio & 76.2 (+3.8) \\
Last token $\oplus$ raw audio & 76.2 (+3.8) \\
First/mid/last $\oplus$ projected audio & 76.6 (+4.2) \\
\midrule
\multicolumn{2}{c}{\textit{LLM hidden states $\oplus$ audio (cross-attention)}} \\
Last token $\oplus$ audio (cross-attn) & 76.4 (+4.0) \\
Last-mean $\oplus$ audio (cross-attn) & 72.4 (0.0) \\
\bottomrule
\end{tabular}
\end{table}

\subsubsection{Regularization}\label{sec:regularization}

Overfitting is the primary bottleneck to approaching oracle performance. Because the selector is trained on a small dataset (${\approx}$7{,}500 examples) relative to a much higher-dimensional input, training and validation accuracy diverge rapidly without regularization. We explored an extensive array of strategies, including weight decay, mixup, feature dropout, input dropout, feature noise, and label smoothing. The unregularized baseline reaches 75.6\% test accuracy before early stopping fires around epoch 25.

Label smoothing is effective because hard oracle targets create a sharp discontinuity at the yes/no decision boundary, where nearly identical branch outputs can receive opposite 0/1 labels. Complementary strategies such as feature noise further reduce the risk of latching onto spurious feature dimensions. The optimal configuration (label smoothing $\epsilon{=}0.25$, feature noise $0.10$, input dropout $0.05$) extends training to 42--75 epochs across splits and improves peak test accuracy to \textbf{76.7\%}.

\subsubsection{Other Tasks}\label{sec:other_tasks}

When the baseline is near chance, contrastive decoding offers little traction and the selector has no useful signal to learn from. On AH Attribute, both Qwen2 (51.0\%) and AF3 (56.0\%) remain near chance regardless of perturbation, consistent with prior benchmarking~\cite{kuan2025can} identifying this as a fundamental capability limit. The same applies to Qwen2 on AH Order, which barely exceeds chance across all branches.

AF3 on AH Order presents a different failure mode: the model is capable (74.7\% baseline), yet the top perturbations are all temporal disruptions---reverse, spectral reverse, and segment shuffle---that heavily overlap in which examples they correct. Reverse dominates at 81.4\%, leaving the selector little signal to improve on the fixed best branch.

On Clotho-AQA, the selector underperforms the best fixed branch as soon as more than one candidate perturbation is considered.

\section{Conclusion}

We demonstrate that expanding inference-time contrastive decoding beyond static baselines improves hallucination suppression. The effectiveness of negative branches varies substantially across tasks and examples, making static perturbations inherently limited. Our results show that LALM hidden states provide useful routing signals for adaptive perturbation selection, while the remaining oracle gap leaves substantial room for improvement.

The remaining oracle gap defines a clear research agenda:
\begin{itemize}
  \item \textit{Scaling the Routing Signal.} Closing the oracle gap requires more training examples in which different perturbations are uniquely useful for distinct subsets of inputs. Larger, more diverse datasets designed to stress-test specific acoustic reasoning skills would provide the routing signal needed to improve selector performance.
  \item \textit{Intrinsic Perturbation Awareness.} Currently, the selector relies on LALM hidden states that are unexposed to our perturbation library. Fine-tuning the LALM using contrastive objectives to explicitly predict branch utility would force its representations to natively encode these acoustic failure modes. As demonstrated by mDPO~\cite{wang2024mdpoconditionalpreferenceoptimization}, models can successfully internalize corrections derived from such multimodal perturbations.
  \item \textit{Contrastive Decoding for Open-ended Questions.} While our oracle framework extends naturally to multiple-choice settings, the next major frontier is open-answer generation and captioning. Expanding adaptive CD from a single binary token to full sequence-level divergence will test the ultimate limits of perturbation-based grounding.
\end{itemize}

\section*{Acknowledgments}
Weiran Wang is supported by a Google Gift Award.

Parts of the experimental codebase were implemented and parts of this manuscript were drafted with assistance from Claude and Claude Code (Anthropic). The entire codebase and the paper were reviewed and edited by the authors.

\bibliographystyle{IEEEtran}
\bibliography{control,references}

\end{document}